\documentclass[aps,prl,twocolumn,showpacs,amsmath,amsmath,amssymb,amsfonts,superscriptaddress]{revtex4-1}
\usepackage{amssymb}
\usepackage{graphicx}% Include figure files
\usepackage{dcolumn}% Align table columns on decimal point
\usepackage{bm}% bold math
\usepackage{psfrag}
\usepackage{bbold}
\usepackage{float}
\usepackage{amsmath}

\begin{document}
\author{B. Olmos}
\affiliation{School
of Physics and Astronomy, The University of Nottingham, Nottingham,
NG7 2RD, United Kingdom}
\author{D. Yu}
\affiliation{School
of Physics and Astronomy, The University of Nottingham, Nottingham,
NG7 2RD, United Kingdom}
\affiliation{Department of Applied Physics, Graduate School of Engineering, The University of Tokyo, Bunkyo-ku, Tokyo 113-8656, Japan}
\author{Y. Singh}
\affiliation{School
of Physics and Astronomy, The University of Birmingham, Birmingham,
B15 2TT, United Kingdom}
\author{F. Schreck}
\affiliation{Institut f\"ur Quantenoptik und Quanteninformation (IQOQI), \"Osterreichische Akademie der Wissenschaften, 6020 Innsbruck, Austria}
\author{K. Bongs}
\affiliation{School
of Physics and Astronomy, The University of Birmingham, Birmingham,
B15 2TT, United Kingdom}
\author{I. Lesanovsky}
\affiliation{School
of Physics and Astronomy, The University of Nottingham, Nottingham,
NG7 2RD, United Kingdom}
\title{Long-range interacting many-body systems with alkaline-earth-metal atoms}
\date{\today}
\begin{abstract}
Alkaline-earth-metal atoms can exhibit long-range dipolar interactions, which are generated via the coherent exchange of photons on the $^3$P$_0-^3$D$_1$-transition of the triplet manifold. In case of bosonic strontium, which we discuss here, this transition has a wavelength of $2.6$ $\mu$m and a dipole moment of $4.03$ Debye, and there exists a magic wavelength permitting the creation of optical lattices that are identical for the states $^3$P$_0$ and $^3$D$_1$. This interaction enables the realization and study of mixtures of hard-core lattice bosons featuring long-range hopping, with tuneable disorder and anisotropy. We derive the many-body Master equation, investigate the dynamics of excitation transport and analyze spectroscopic signatures stemming from coherent long-range interactions and collective dissipation. Our results show that lattice gases of alkaline-earth-metal atoms permit the creation of long-lived collective atomic states and constitute a simple and versatile platform for the exploration of many-body systems with long-range interactions. As such, they represent an alternative to current related efforts employing Rydberg gases, atoms with large magnetic moment, or polar molecules.
\end{abstract}


\maketitle
\emph{Introduction.-} The creation and exploration of quantum systems with long-range interactions is in the focus of intense research activity worldwide. Within the context of novel technological applications, such as quantum information processing, strong long-range interactions are essential as they permit the implementation of entangling gate operations among distant qubits. From the perspective of fundamental physics of condensed matter systems, these interactions permit the study of strongly correlated phases of quantum matter. In order to access this potential there is a need for a simple experimental platform that fosters long-range interactions. In the domain of ultra cold gases there are currently three approaches, which rely on atoms with large magnetic dipole moment (e.g. chromium, dysprosium and erbium \cite{Griesmaier05,Lu11,Aikawa12}), polar molecules \cite{Ni08} or atoms excited to Rydberg states. In particular Rydberg atoms have celebrated recent successes in the realms of fundamental physics and technological applications \cite{Saffman10-2,Weimer10-2,Schachenmayer10,Lesanovsky11}. Two experiments have successfully implemented quantum gate protocols among qubits encoded in distant atoms \cite{Urban09,Gaetan09} and, very recently, the intricate dynamics of strongly correlated Rydberg lattice gases was studied in experiment \cite{Viteau11,Anderson11,Schauss12}.

\begin{figure}
  \includegraphics[width=\columnwidth]{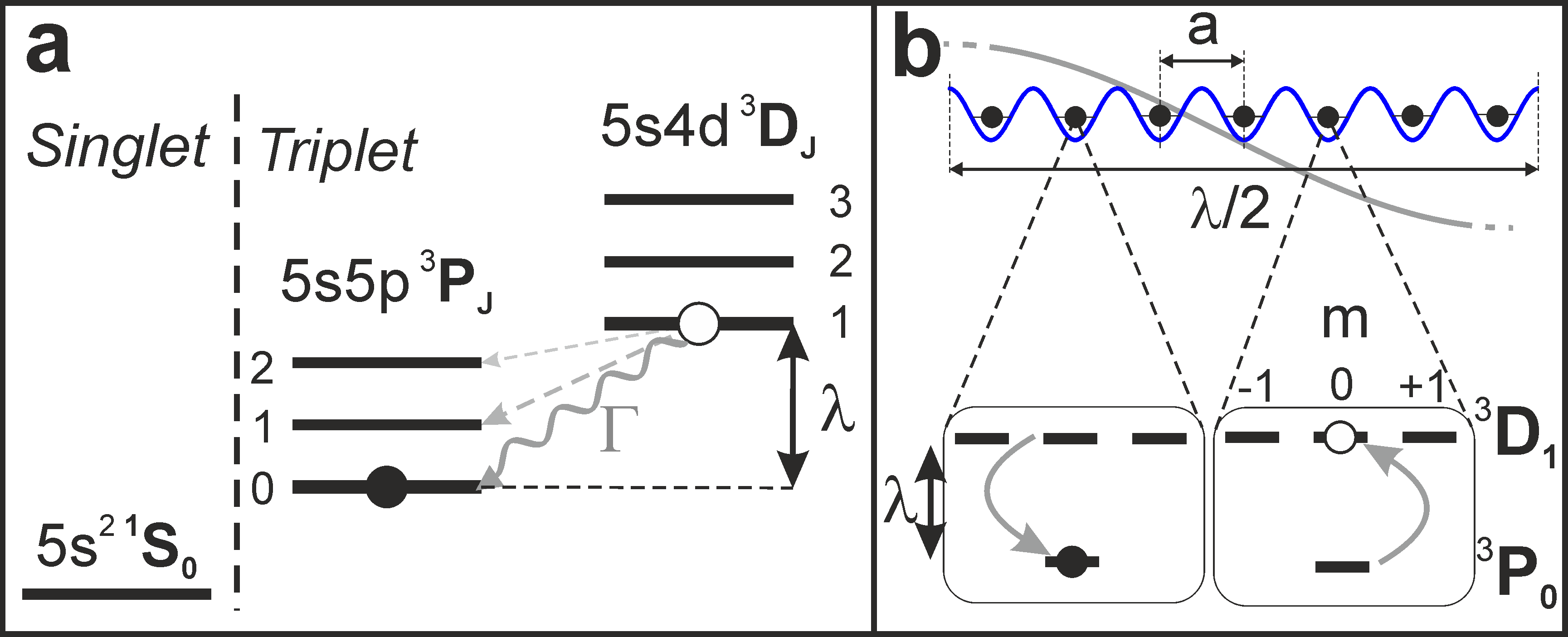}
  \caption{\textbf{a}: Relevant levels of the Sr atom. \textbf{b}: Atoms are trapped in an optical lattice and the interaction between them is generated by the exchange of (virtual) photons on the transition between $^3$P$_0$ and the three degenerate $^3$D$_1$ states. The decay rate and wavelength are $\Gamma=290\times10^3$ s$^{-1}$ and  $\lambda=2.6$ $\mu$m, respectively. $\lambda$ is much larger than the typical interatomic spacing ($a=206.4$ nm) at the magic wavelength.}\label{fig:levels}
\end{figure}
In this work we describe a novel platform for the realization of many-body systems featuring long-range dipolar interactions. It is based on the exchange of virtual photons between low-lying triplet states of alkaline-earth-metal atoms, building on the seminal work by Brennen et al. \cite{Brennen99}. The interaction strength can be comparable to the one typically achievable with polar molecules, i.e. three orders of magnitude stronger than among atoms with large magnetic dipole moments. Compared to Rydberg atoms the interactions are substantially weaker. However, the use of low-lying states makes our system less prone to perturbing electric fields and reduces the number of radiative decay channels and involved levels. This might offer an interesting perspective for the study of open quantum spin systems.

We specifically focus on bosonic strontium (Sr) atoms trapped in a deep optical lattice in a Mott insulator state and photons exchanged on the $^3$P$_0-^3$D$_1$-transition (see Fig. \ref{fig:levels}a). We derive the corresponding many-body Master equation, focussing specifically on the situation of planar and linear $xy$-models with long-range interactions, which are equivalent to hard-core lattice bosons with long-range hopping. We provide data of a magic wavelength for an optical lattice that grants equal confinement for both the states $^3$P$_0$ and $^3$D$_1$, characterize the role of decoherence and disorder and show how the interactions and the resulting collective light-scattering become manifest in the fluorescence spectrum. Building on the routine creation of alkaline-earth-metal Mott insulators in a number of laboratories \cite{Fukuhara09,Stellmer12}, our approach represents a simple route for the exploration and exploitation of many-body phenomena in long-range interacting systems and highlights a novel way for the creation of long-lived collective atomic states, with applications in quantum optics and quantum information.

\emph{Triplet states of strontium.-} Sr has two valence electrons and its spectrum is thus formed by a series of singlet and triplet states (Fig. \ref{fig:levels}a). The radiative transitions between the two series are dipole forbidden, which - due to the resulting small transition line widths - leads to a wide range of applications, such as ultra precise atomic clocks \cite{Takamoto05,*Ye08,*Akatsuka08} or the implementation of quantum information processing protocols \cite{Saffman10-2,Gorshkov10,*Gorshkov09,*Daley08}. Here, we consider Sr atoms in a Mott insulator state \cite{Fukuhara09,Stellmer12} as depicted in Fig. \ref{fig:levels}b. The lattice is identical for the $^3$P$_0$ and $^3$D$_1$ states, and its blue-detuned magic wavelength is located at $\lambda_{bm}=412.8$ nm (for more detailed information, see Supplemental Material). The resulting lattice spacing is $a=206.4$ nm, a value that we will use throughout this work to benchmark our results.

Initially all Sr atoms within the lattice are excited to the triplet manifold, i.e. either to the metastable state $^3$P$_0$ (its lifetime can be considered infinite for all experimental purposes) or the state $^3$D$_1$. Long-range interactions between two Sr atoms then emerge by the resonant exchange of photons that are emitted on the transition $^3$D$_1-^3$P$_0$ which has the wavelength $\lambda=2.6$ $\mu$m (see Fig. \ref{fig:levels}b). This mechanism, which leads to a dipolar interaction, is in general well-understood \cite{Dicke54,Agarwal70,Lehmberg70} but usually it plays a role only in very dense samples, i.e. where the interatomic distance is far smaller than the wavelength of the transition \cite{Zoubi10,*Zoubi11,*Zoubi11-2,Brennen99,Keaveney12}. Conventional lattice setups usually do not reach such parameters. However, the combination of a lattice with small spacing and a long wavelength transition that is available in Sr allows us to enter this regime without having to deal with the destructive effect of atomic collisions. The transition dipole moment between the $^3$P$_0$ state and the three degenerate $^3$D$_1$-states is $p=4.03$ Debye and effectuates a strong resonant dipole-dipole interaction that extends over several lattice sites as depicted in Fig. \ref{fig:levels}b. In order to avoid unnecessary complications we restrict ourselves to photon exchange on the $^3$D$_1-^3$P$_0$ transition, which has the highest decay rate ($\Gamma=290\times10^3$ s$^{-1}$, \cite{Zhou10}) and leave the (straightforward) consideration of the additional weaker coupling channels $^3$D$_1-^3$P$_1$ and $^3$D$_1-^3$P$_2$ to future investigations.

\emph{Many-body Master equation.-} The starting point for the derivation of the many-body Master equation is the Hamiltonian describing $N$ Sr atoms coupled to the radiation field. To formulate it we introduce the vector transition operator for the $k$-th atom (located at $\mathbf{r}_k$) $\mathbf{b}_k=b_{kx}\,\hat{x}+b_{ky}\,\hat{y}+b_{kz}\,\hat{z}$ such that the transition dipole matrix elements are real and aligned along the three cartesian spatial axes $\hat{x},\hat{y}$ and $\hat{z}$ \cite{James93}. Here, $b_{kj}=\left|P\right>_k\left<j\right|$ with $j=x,y,z$, where $\left|P\right>_k$ represents the $k$-th atom in the $^3$P$_0$ state and $\left|j\right>_k$ the cartesian states of $^3$D$_1$, related to the angular momentum ones $\left|m\right>_k$ (with $m=-1,0,+1$) as $\left|\mp1\right>= \left(\pm\left|x\right>-i\left|y\right>\right)/\sqrt{2}$ and $\left|0\right>=\left|z\right>$. Within this notation the Hamiltonian of the atomic ensemble and the radiation field is given by $H_\mathrm{af}=\sum_{k=1}^N\hbar\omega_\mathrm{a}\mathbf{b}_k^\dagger\cdot\mathbf{b}_k +\sum_{\mathbf{q}\lambda}\hbar\omega_{\mathbf{q}}a_{\mathbf{q}\lambda}^\dagger a_{\mathbf{q}\lambda}+i\hbar\sum_{k=1}^N\sum_{\mathbf{q}\lambda}
\mathbf{g}_{\mathbf{q}\lambda}\cdot\left(a_{\mathbf{q}\lambda}^\dagger
\mathbf{s}_k e^{-i\mathbf{q}\cdot\mathbf{r}_k}-
\mathbf{s}_ka_{\mathbf{q}\lambda}e^{i\mathbf{q}\cdot\mathbf{r}_k}\right)$ with $\mathbf{s}_k=\mathbf{b}_k^\dag+\mathbf{b}_k$, where $\hbar\omega_\mathrm{a}=2\pi\hbar c/\lambda$ is the energy difference between the $^3$P$_0$ state and the degenerate $^3$D$_1$-manifold, $\hbar\omega_{\mathbf{q}}$ is the energy of a photon with momentum $\mathbf{q}$ and polarization $\lambda$ and $a_{\mathbf{q}\lambda}$ is the annihilation operator of such a photon (bosonic operators, i.e., $[a_{\mathbf{q}\lambda},a_{\mathbf{q}'\lambda'}^\dag] =\delta_{\mathbf{q}\mathbf{q}'}\delta_{\lambda\lambda'}$). The coefficient $\mathbf{g}_{\mathbf{q}\lambda}$ is given by
$\mathbf{g}_{\mathbf{q}\lambda}=p\sqrt{\frac{\omega_{\mathbf{q}}}{2\epsilon_0\hbar
V}} \hat{e}_{\mathbf{q}\lambda}$,
with $V$ the quantization volume, and $\hat{e}_{\mathbf{q}\lambda}$ the unit polarization vector of the photon ($\mathbf{q}\cdot\hat{e}_{\mathbf{q}\lambda}=0$).

Following Refs. \cite{Lehmberg70,James93} we obtain the Master equation governing the evolution of the density matrix $\rho$ of the atomic ensemble: $\dot{\rho}=-\frac{i}{\hbar}\left[H,\rho\right]+{\cal D}(\rho)$. The first term depends on the many-body Hamiltonian
\begin{equation}
  H=\hbar\omega_\mathrm{a}\sum_{k}\mathbf{b}_k^\dagger\cdot\mathbf{b}_k+\hbar\sum_{k\neq
l}\mathbf{b}^\dag_{k}\cdot\overline{V_{kl}}\cdot\mathbf{b}_{l}. \label{eq:hamiltonian}
\end{equation}
Its first part contains the bare energies of the atomic levels and the second part describes the long-ranged and (in general) anisotropic dipole-dipole interaction, characterized by the coefficient matrix
\begin{equation*}
  \overline{V_{kl}}=\frac{3\Gamma}{4}\left\{\left[y_0\left(\kappa_{kl}\right)-
\frac{y_1\left(\kappa_{kl}\right)}{\kappa_{kl}}\right]\mathbb{1}+
y_2\left(\kappa_{kl}\right)\hat{r}_{kl}\hat{r}_{kl}\right\}.
\end{equation*}
Here $y_n(x)$ represents the $n$-th order spherical Bessel
function of the second kind and $\kappa_{kl}\equiv k_\mathrm{a}r_{kl}$ with $k_\mathrm{a}=\omega_\mathrm{a}/c$ and $\mathbf{r}_{kl}=\mathbf{r}_{k}-\mathbf{r}_{l}=r_{kl}\hat{r}_{kl}$.
The second term of the Master equation depends on the dissipator
\begin{equation*}
{\cal D}(\rho)=\sum_{kl}\mathbf{b}_{k}\cdot\overline{\Gamma_{kl}}\cdot \rho\mathbf{b}^\dag_{l}
-\frac{1}{2}\left\{\mathbf{b}_k^\dag\cdot\overline{\Gamma_{kl}}\cdot\mathbf{b}_{l},\rho\right\}.
\end{equation*}
The coefficient matrix
\begin{eqnarray*}
\overline{\Gamma_{kl}}=\frac{3\Gamma}{2}\left\{\left[j_0\left(\kappa_{kl}\right)-
\frac{j_1\left(\kappa_{kl}\right)}{\kappa_{kl}}\right]\mathbb{1}+ j_2\left(\kappa_{kl}\right)\hat{r}_{kl}\hat{r}_{kl}\right\},
\end{eqnarray*}
encodes the dissipative couplings among the atoms and $j_n(x)$ represents the $n$-th order spherical Bessel function of the first kind.

\begin{figure}
  \includegraphics[width=\columnwidth]{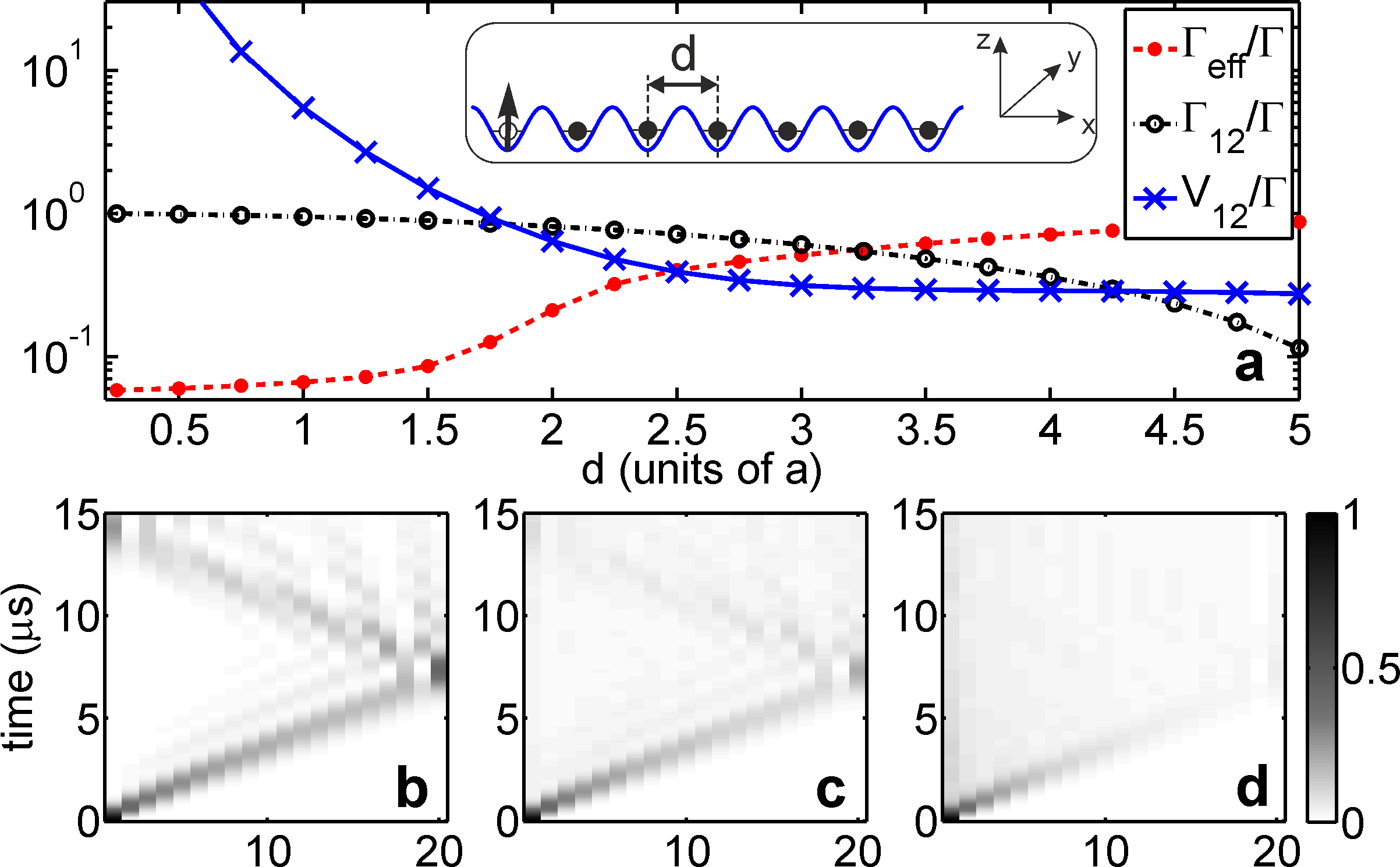}
  \caption{Transport of a single excitation on a chain of $N=20$ atoms with induced dipoles pointing in the $z$-direction. \textbf{a:} Comparison between the nearest neighbor interaction $V_{12}$, the damping rate $\Gamma_{12}$, and the effective decay rate $\Gamma_\mathrm{eff}$ of the single excitation for different values of the nearest neighbor separation $d$. For $d=a$ the effective decay rate is approximately two orders of magnitude smaller than the coherent interaction. \textbf{b, c, d:} Time-evolution of a single, initially localized excitation, for varying disorder and for $d=a$ with: \textbf{b}: $\sigma/a=0$, \textbf{c}: $\sigma/a=0.025$ and \textbf{d}: $\sigma/a=0.05$.}\label{fig:transport}
\end{figure}

The coherent and dissipative dynamics are intimately connected since they both originate from the emission/absorption of photons. However, by virtue of the long wavelength of the photons and the achievable small interparticle separation (see Fig. \ref{fig:levels}b) one can reach a parameter regime in which the coherent interaction is much stronger than the dissipation. This is shown in Fig. \ref{fig:transport}a where we compare the coherent interaction $V_{12}$ to the dissipative rate $\Gamma_{12}$ for two atoms separated by a distance $d$ whose induced dipoles are aligned as $\uparrow\uparrow$. For $d=a$ the ratio $V_{12}/\Gamma_{12}$ is approximately $5.7$, but for the dipole alignment $\rightarrow\rightarrow$ this ratio reaches $13.9$, showing that the Sr lattice setup is indeed well-suited for the study of coherent many-body phenomena. In fact, we will later see that due to the presence of sub-radiant states even larger ratios can be achieved.

\emph{Coherent dynamics.-} Hamiltonian (\ref{eq:hamiltonian}) conserves the number of atoms in the $^3$D$_1$-state, $N_\mathrm{D}=\sum_{k} \mathbf{b}_k^\dagger\cdot\mathbf{b}_k$. Hence, non-trivial dynamics is only induced by the second term of eq. (\ref{eq:hamiltonian}) which depends strongly on the geometry of the lattice. To get a glimpse of the versatility of the Sr setup for the study of coherent many-body phenomena let us consider a situation in which we have a one- or two-dimensional lattice located in the $x-y$ plane and where atoms are solely excited to the $^3$D$_1(m=0)$-state. In this case the Hamiltonian (\ref{eq:hamiltonian}) simplifies to $H_{xy}=\sum_{k\neq l} W_{kl}b_{kz}^\dag b_{lz}$ with the interaction coefficients $W_{kl}=\frac{3\hbar\Gamma}{4}\left[y_0\left(\kappa_{kl}\right)-
\frac{y_1\left(\kappa_{kl}\right)}{\kappa_{kl}}\right]\approx \frac{3\hbar\Gamma}{4 k^3_\mathrm{a}}\times \frac{1}{|\mathbf{r}_k-\mathbf{r}_l|^3}$ (the dipolar approximation holds when $\kappa_{kl}\lesssim 1$). $H_{xy}$ represents an $xy$-model with long-range interaction or equivalently a system of hard-core bosons with long-range hopping, which is due to the fact that the operators $b_{kz}^\dag$ and $b_{kz}$ can be identified with creation/annihilation operators of hard-core bosons. Note, that such long-range hopping can also be effectively established in ion traps \cite{Deng08,Hauke10}. For nearest neighbor interactions the $xy$-model has been studied extensively in the literature in the context of quantum information processing \cite{Wang01}, quantum and thermal phase transitions \cite{Kennedy88,*Kubo88,*Fradkin89,*Harada98,*Sandvik99} and relaxation of closed quantum systems \cite{Rigol08}. The case of spin-systems with long-range interactions is less explored \cite{Fisher72,*Sak73,*Kosterlitz76} and recent Monte Carlo simulations \cite{Picco12} raise new questions concerning their phase behavior.

To study the (thermo)dynamics of $H_{xy}$ one needs to control the density of hard-core bosons $N_\mathrm{D}/N$ which is done as follows: Starting with all atoms in the state $^3$P$_0$ one irradiates a laser on the $^3$P$_0$-$^3$D$_1$-transition. The Hamiltonian describing the atom-laser coupling is $H_\mathrm{L}=\hbar\mathbf{\Omega}_\mathrm{L}\cdot\sum_{k=1}^N\left(e^{-i\left(\mathbf{k}\cdot\mathbf{r}_k-\omega t\right)}\mathbf{b}_k +e^{i\left(\mathbf{k}\cdot\mathbf{r}_k-\omega t\right)}\mathbf{b}^\dag_k\right)$ with $\mathbf{\Omega}_\mathrm{L}=p\mathbf{E}_0/\hbar$, where  $\omega$ is the frequency, $\mathbf{k}$ the momentum and $\mathbf{E}_0$ the amplitude vector of the laser. Applying a strong laser pulse ($|\mathbf{\Omega}_\mathrm{L}|\gg |W_{kl}|$) with amplitude vector $\mathbf{E}_0=E_0 \hat{z}$ for a time $\tau$ on resonance ($\omega=\omega_\mathrm{a}$) creates on average $N_\mathrm{D}(\tau)=N\sin^2(2|\mathbf{\Omega}_\mathrm{L}|\tau)$ hard core-bosons. This number fluctuates with standard deviation $\Delta N_\mathrm{D} = \sqrt{N}|\sin(4|\mathbf{\Omega}_\mathrm{L}|\tau)|/2$. Alternatively, one could think of selectively changing the state of atoms in specific sites. In particular, in the one-dimensional case, the required single-site addressing can be achieved by applying a magnetic field gradient, which is switched off after the desired state is prepared \cite{Bakr09,Sherson10}.

The $xy$-model focussed on here merely represents a very simple scenario. In all its generality the Hamiltonian (\ref{eq:hamiltonian}) describes three species of hard-core bosons which, depending on the lattice dimension and geometry, exhibit anisotropic and species-dependent long-range hopping and species interconversion. The density of the individual species is furthermore controllable by simple laser pulses. This generates a rich playground for the discovery and study of many-body quantum phases.

\emph{Dissipation and disorder.-} We now discuss two seemingly harmful effects: Collective dissipation due to radiative decay and disorder stemming from the uncertainty in the atomic positions. To analyze them we numerically simulate a situation in which a single excitation (or hard-core boson) propagates through a linear chain of $N=20$ Sr atoms orientated along the $x$-direction under the action of $H_{xy}$ and the corresponding dissipator. Initially, the excitation is localized at the leftmost atom which is in the $^3$D$_1(m=0)$-state. Its propagation is depicted in Fig. \ref{fig:transport}b which also shows an expected overall decrease of the excitation density due to dissipation. Moreover, we have extracted from the simulations the (effective) decay rate $\Gamma_\mathrm{eff}$ of the excitation as a function of the lattice spacing $d$ (see Fig. \ref{fig:transport}a). Surprisingly, one can see that for $d=a$ the effective decay rate of the excitation is not only much smaller than the single-atom one $\Gamma$, but more than two orders of magnitude smaller than the dipole-dipole interaction. The reason for this unexpected long lifetime of the excitation resides in the collective character of the dissipation and the emergence of sub-radiant states.

Let us now discuss disorder, which arises from the fact that the external wavefunction of each localized Sr atom has a finite width. We model this wavefunction as a three-dimensional isotropic Gaussian with width $\sigma$, localized on the respective lattice site. Using the dipolar approximation for the interaction and assuming $r_{kl}\gg \sigma$ the couplings $W_{kl}$ become random variables which are distributed according to
$p(W_{kl})=[(A^{2/3})/(3\sqrt{2\pi}\sigma  r_{kl}W_{kl}^{5/3}]\exp[-(r_{kl}- [A/W_{kl}]^{1/3})^2/(2\sigma^2)]$,
where $A=3\hbar\Gamma/(4 k^3_\mathrm{a})$. Again, the simulated excitation transport reveals the effect of disorder in Figs. \ref{fig:transport}b,c and d for $d=a$. With increasing ratio $\sigma/a$ transport becomes less efficient and more population remains localized at the left of the lattice. Beyond this simple illustration it will be very interesting to study the effect of this controllable disorder in the many-body context. It has been shown that the ground state of hard-core bosons exhibits a localization transition for certain types of disorder \cite{Pazmandi95} in the hopping rates. It is an open question whether this transition is present also here.

\begin{figure}
  \includegraphics[width=\columnwidth]{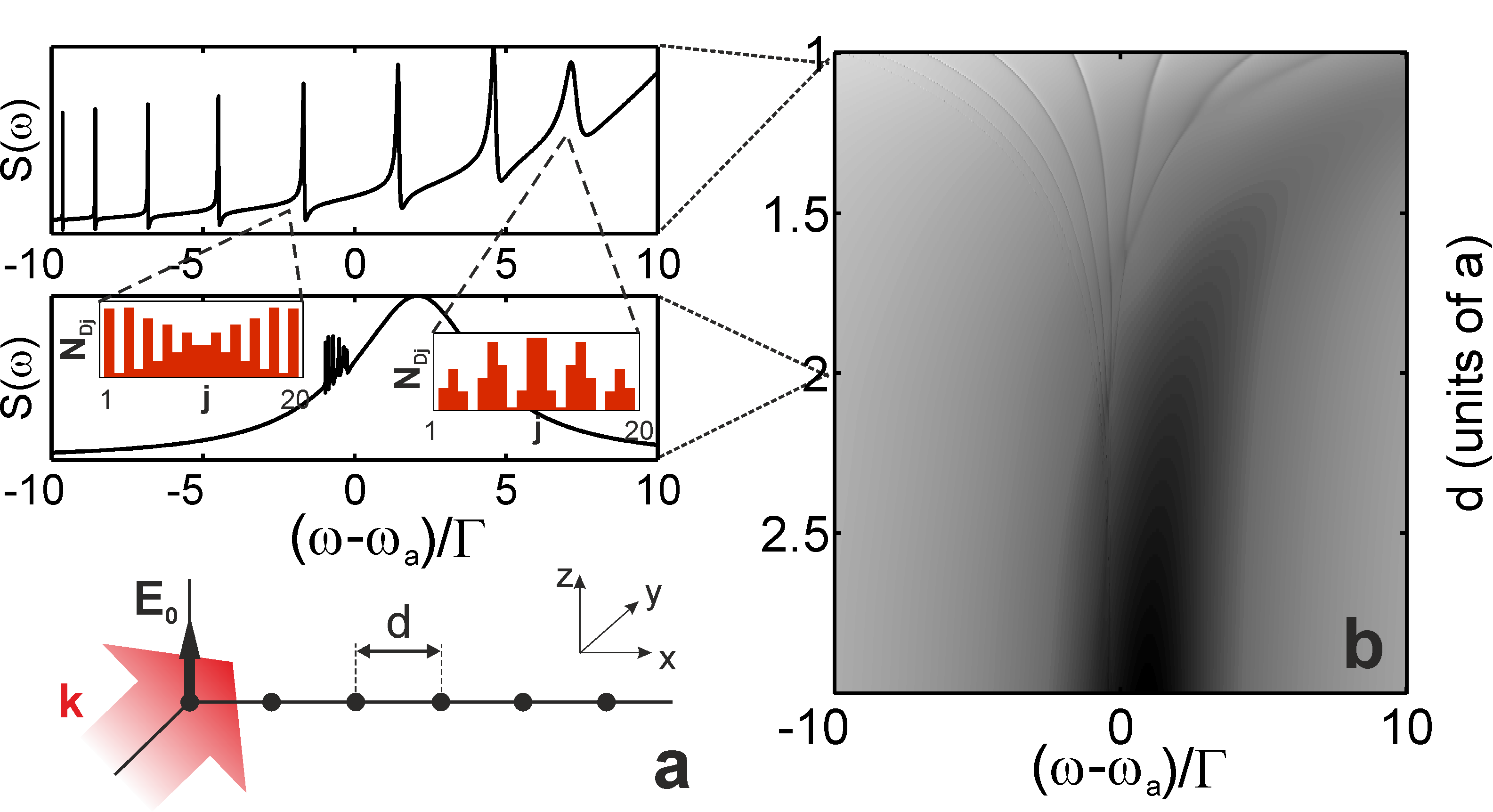}
  \caption{\textbf{a}: One-dimensional lattice (oriented along the $x$ axis) with lattice constant $d$. A laser is shone on the atoms with momentum $\mathbf{k}$ and polarization $\mathbf{E}_0$ both perpendicular to the chain. \textbf{b}: Power spectrum of the radiation scattered into the $y-z$ plane $S(\omega)$ (arbitrary units and logarithmic scale) as a function of the interparticle separation $d$ and the frequency of the laser $\omega$. The two insets show the corresponding spectrum for $d=a$ and $d=2a$ (linear scale). The average excitation number on the $j$-th lattice site $N_{\mathrm{D},j}$, corresponding to the states represented by two of the peaks is also sketched.}\label{fig:spectrum}
\end{figure}

\emph{Spectroscopy on the few-body system.-} Finally, let us discuss the spectroscopic properties of the Sr lattice system and find out whether they provide a clear experimental signature of the presence of long-range interactions. To this end we calculate the power spectrum of the radiation field $\mathbf{E}(\mathbf{R},t)$ scattered off the ensemble of atoms in the direction of observation $\mathbf{R}$ driven by a weak incident laser field of frequency $\omega$: $S(\mathbf{R},\omega)=\frac{1}{2\pi}\int_{-\infty}^{\infty}d\tau e^{i\omega\tau}\left<\mathbf{E}^\dag(\mathbf{R},t)\cdot \mathbf{E}(\mathbf{R},t+\tau)\right>$. We consider again the above-discussed case of a one-dimensional lattice with $N$ sites oriented along the $x$-axis where nearest neighbors are separated by the distance $d$. The dynamics of this driven ensemble is described by Hamiltonian $H_{xy}$, the corresponding dissipator and the Hamiltonian $H_\mathrm{L}$, which takes into account the action of the laser field with polarization $\mathbf{E}_0=E_0 \hat{z}$. For the sake of simplicity we orient the laser beam such that its momentum $\mathbf{k}$ is perpendicular to the chain (see sketch in Fig. \ref{fig:spectrum}a) and observe the spectrum of the radiation scattered into the $y-z$ plane (denoted by $S(\omega)$).

Before discussing the results for a chain, let us first consider the case of two atoms, i.e. $N=2$, where the problem can be treated analytically. Here, the spectrum of the radiation $S(\omega)$ consists of a single peak of Lorentzian form $S(\omega)\propto1/[(\omega-\omega_\mathrm{a}-W_{12}/\hbar)^2 +(\Gamma+\Gamma_{12})^2/4]$
with $\Gamma_{12}=\frac{3\Gamma}{2}[j_0\left(kd\right)-
\frac{j_1\left(kd\right)}{kd}]$. Hence, the signature of the strong interaction is a shift towards the blue (repulsive interaction) and broadening of the Lorentzian peak \cite{Das08,Wang10}. Note that only one peak (the symmetric superposition of the two singly-excited states) appears in this case. This is due to two facts: (i) In this particular geometry (the momentum of the laser being perpendicular to the interparticle separation) the laser only couples to the (super-radiant) symmetric state and (ii) the symmetric and anti-symmetric states are eigenstates of the dissipator as well as of the Hamiltonian and, hence, the dissipation induces no couplings between the symmetric and anti-symmetric states.

For a chain of $N=20$ atoms we have calculated the spectrum numerically and the result is shown in Fig. \ref{fig:spectrum}b. One observes that, as in the case of two atoms, the laser field couples to the symmetric (spin wave) state, which decays super-radiantly (visible as the very broad feature on the blue side of the atomic line). More interesting, however, is the emergence of a number of narrow peaks belonging to long-lived sub-radiant states \cite{Tokihiro93}. These lines appear due to the fact that the coherent and dissipative interaction do not share the same set of eigenstates  \footnote{Note, that in the limit $N\rightarrow\infty$ or in case of periodic boundary conditions the two sets of eigenstates coincide and thus these narrow lines are absent \cite{Nienhuis87}.}. This admixes a fraction of the symmetric state to other collective states, making them excitable by the laser. The density profile of two selected states is shown in the inset of Fig. \ref{fig:spectrum}b. The Sr system thus offers the possibility of exciting long-lived collective states \cite{Jenkins12,Wiegner11} which can find an application in quantum information and photon storage.

\emph{Conclusions and outlook.-} In conclusion, we have shown that alkaline-earth-metal atoms in optical lattices provide a platform offering controllable many-body systems with long-range interactions. Specifically, we have analyzed a regime which implements hard-core bosons with coherent long-range hopping. Beyond that simple example, the system permits the study of mixtures of hard-core bosons in the presence of tunable disorder. Signatures of the long-range interaction are manifest in the spectrum of the radiation which is collectively scattered from the atomic ensemble. In the future we will investigate closer the properties of the emitted light \cite{Olmos10} and its use to detect quantum phases \cite{Mekhov07} and (disorder-driven) phase transitions. We will furthermore investigate dynamical phases \cite{Lesanovsky12b} resulting from the interplay between dissipative and coherent dynamics and explore how the versatility of the Sr platform can be further enhanced, e.g., by the application of static electric and microwave fields.

\acknowledgments
\emph{Acknowledgments.-} K.B. acknowledges support under EPSRC Grant No. EP/E036473/1. Y.S. acknowledges support under EC Grant No. 255000. I.L. acknowledges funding by EPSRC and the EU FET-Young Explorers QuILMI and gratefully acknowledges discussions with A. Trombettoni. B.O. acknowledges funding by the University of Nottingham. F.S. acknowledges support by the Austrian Ministry of Science and Research (BMWF) and the Austrian Science Fund (FWF) through a START grant under Project No. Y507-N20. K.B. and F.S. acknowledge support under EC FET-Open Grant No. 250072.

\appendix

\setcounter{figure}{0}
\setcounter{table}{0}
\setcounter{equation}{0}
\renewcommand{\thefigure}{S\arabic{figure}}
\renewcommand{\thetable}{S\Roman{table}}
\renewcommand{\theequation}{S\arabic{equation}}

\section{Supplemental Material for "Long-range interacting many-body systems with alkaline-earth-metal atoms"}

In this supplementary material, we provide some details on a possible implementation of an optical lattice configuration for strontium atoms, which grants equal confinement for the $(5s5p)^{3}$P$_{0}$ state and \emph{all} magnetic sub-levels of the $(5s4d)^{3}$D$_{1}$ state, thereby realizing a low-dimensional long-range interacting many-body system.

We consider a one-dimensional (1D) blue-detuned optical lattice along the $x$-axis. The positive frequency part of the lattice field is given by
$\textbf{E}^{(+)}_{b}(x)e^{-i\omega_{b}t}={\cal{E}}_{b}\left(\sqrt{\frac{2}{3}}\hat{y}+\sqrt{\frac{1}{3}}\hat{z}\right)\cos\left(\frac{2\pi
x}{\lambda_{b}}\right)e^{-i\omega_{b}t}$. Here, ${\cal{E}}_{b}$ is the laser field amplitude, $\omega_{b}=2\pi c/\lambda_{b}$ is the laser frequency and $c$ is the speed of light in vacuum. Following Ref.~\cite{Rosenbusch09}, one finds that with this choice of the lattice field all three states $(5s4d)^{3}$D$_{1}(m=1,0,-1)$ have the same ac polarizability $\alpha^{b}_{D}(\lambda_{b})$.

\begin{figure}
\includegraphics[width=4.3cm]{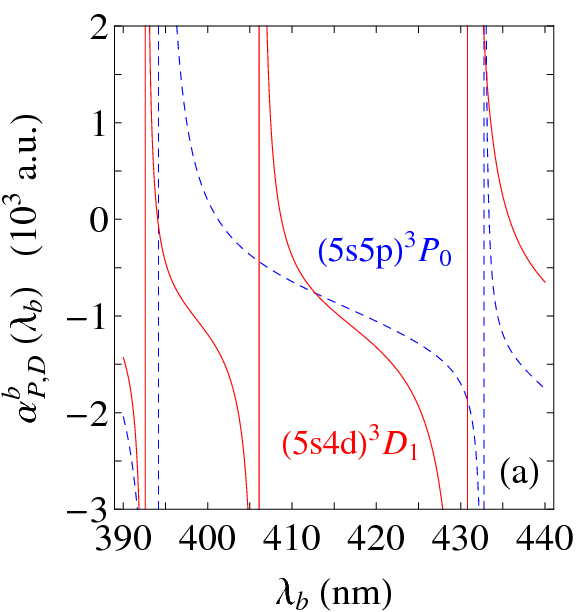}~\includegraphics[width=4.33cm]{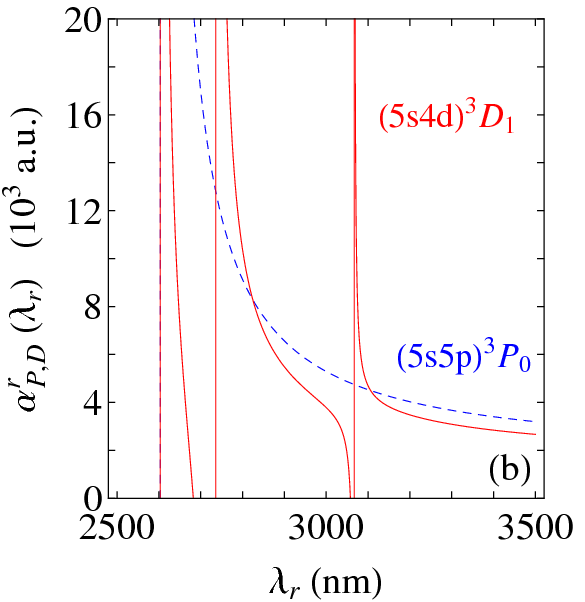}\\
\caption{Wavelength dependence of the ac polarizabilities
$\alpha^{b,r}_{D}(\lambda_{b,r})$ (solid lines) and
$\alpha^{b,r}_{P}(\lambda_{b,r})$ (dashed lines). (a) A magic
wavelength is located on the blue side of the
$(5s5p)^{3}$P$_{0}-(5s4d)^{3}$D$_{1}$ transition. (b) A red-detuned magic wavelength is located between
$(5s5p)^{3}$P$_{1}-(5s4d)^{3}$D$_{1}$ and
$(5s5p)^{3}$P$_{2}-(5s4d)^{3}$D$_{1}$ transitions. In both cases, the ac polarizabilities for all $(5s4d)^{3}$D$_{1}(m=1,0,-1)$ states are same.}\label{Supp1}
\end{figure}

Based on the data for wavelengths and Einstein coefficients for the electric dipole transitions relevant to the $(5s5p)^{3}$P$_{0}$ and $(5s4d)^{3}$D$_{1}$ states in Refs.~\cite{Sansonetti10,Gustavo88,Rubbmark78,Xiaoji10}, we have calculated $\alpha^{b}_{D}(\lambda_{b})$ and also the ac polarizability $\alpha^{b}_{P}(\lambda_{b})$ of the $(5s5p)^{3}$P$_{0}$ state as a function of the lattice wavelength $\lambda_{b}$. In Fig.~\ref{Supp1}(a) we show this data in the vicinity of 400 nm. A magic wavelength is found at $\lambda_{bm}\approx412.8$ nm where $\alpha^{b}_{P}(\lambda_{bm})=\alpha^{b}_{D}(\lambda_{bm})\approx-0.76\times10^{3}$
a.u.

Note that the negative polarizability means that atoms will be trapped in the intensity minima of the lattice field. Thus, in order to tightly trap atoms in the 1D blue-detuned optical lattice in experiment, one needs to provide additional transverse confinement. Without introducing a differential light-shift between the $^3$P$_0$ and $^3$D$_1$ states, this additional confinement can easily be created with a two-dimensional (2D) red-detuned optical lattice near a red magical wavelength. The positive frequency part of the corresponding laser field is given by
$\textbf{E}^{(+)}_{r}(y,z)e^{-i\omega_{r}t}={\cal{E}}_{r}\left[\boldsymbol{\epsilon}_{1}\cos(\textbf{k}_{1}\cdot\textbf{r})+i\boldsymbol{\epsilon}_{2}\cos(\textbf{k}_{2}\cdot\textbf{r})\right]e^{-i\omega_{r}t}$,
where ${\cal{E}}_{r}$ is the lattice field amplitude,
$\boldsymbol{\epsilon}_{1,2}=\sqrt{\frac{2}{3}}\hat{y}\mp\sqrt{\frac{1}{3}}\hat{z}$
are field polarizations and $\textbf{k}_{1,2}=\frac{2\pi}{\lambda_{r}}\left(\sqrt{\frac{1}{3}}\hat{y}\pm\sqrt{\frac{2}{3}}\hat{z}\right)$
and $\omega_{r}=2\pi c/\lambda_{r}$ are the the corresponding wave vectors and frequencies, respectively. Again following Ref.~\cite{Rosenbusch09}, one can prove that all three $(5s4d)^{3}$D$_{1}(m=1,0,-1)$ states exhibit the same
ac polarizability $\alpha^{r}_{D}(\lambda_{r})$. Based on the data in Refs.~\cite{Sansonetti10,Gustavo88,Rubbmark78,Xiaoji10}, we show $\alpha^{r}_{D}(\lambda_{r})$ and the ac polarizability
$\alpha^{r}_{P}(\lambda_{r})$ of the $(5s5p)^{3}$P$_{0}$ state as a function of the wavelength $\lambda_{r}$ in the vicinity of 3 $\mu$m in Fig.~\ref{Supp1}(b). A red-detuned magic wavelength
$\lambda_{rm}\approx2824.6$ nm is located between $(5s5p)^{3}$P$_{1}-(5s4d)^{3}$D$_{1}$ and
$(5s5p)^{3}$P$_{2}-(5s4d)^{3}$D$_{1}$ transitions, at which is
$\alpha^{r}_{P}(\lambda_{rm})=\alpha^{r}_{D}(\lambda_{rm})\approx8.3\times10^{3}$
a.u., and therefore atoms will be trapped in the intensity maxima of the laser field, which permits the straightforward construction of a confining dipole trap.

Note that one can ignore the interference effect between blue- and red-detuned laser fields because of their extremely large frequency difference. Thus, superimposing these two lattice fields creates a setup in which the state $(5s4d)^{3}$P$_{0}$ and all magnetic sub-levels of $(5s4d)^{3}$D$_{1}$ encounter the same trapping potential, and realizes a 1D (along the $x$-axis) long-range interacting many-body system. We would like to point out that this lattice configuration can also be modified to realize a 2D (the $x-y$ plane) long-range interacting many-body system. Note furthermore, that if only one of the magnetic sub-levels of $(5s4d)^{3}$D$_{1}$ is used experimentally (as discussed in the manuscript) one can achieve a magic lattice with less stringent laser field geometries.

\end{document}